\documentclass[twocolumn,aps,pra,superscriptaddress,showpacs]{revtex4-1}
\usepackage{epsfig}
\usepackage[colorlinks=true,citecolor=blue,urlcolor=blue,linkcolor=blue]{hyperref}
\usepackage{chemscheme}
\usepackage{graphicx}
\usepackage{multirow}
\usepackage[table,xcdraw]{xcolor}
\usepackage{natbib}
\usepackage{chemfig}
\usepackage[version=4]{mhchem}
\usepackage[normalem]{ulem}
\usepackage{floatrow}
\usepackage{tabulary}
\usepackage{booktabs}
\usepackage[flushleft]{threeparttable}
\usepackage{multirow}
\usepackage{xspace}
\usepackage{xcolor} 
\definecolor{darkgreen}{rgb}{0.0, 0.6, 0.0}
\DeclareMathAlphabet{\mathbsf}{OT1}{cmss}{bx}{n}

\newcommand{\cm}{cm$^{-1}$\xspace}

\usepackage{xspace}
\newcommand{\pgopher}{\textsc{pgopher}\xspace}
\newcommand{\pKa}{$\mathrm{p}K_\mathrm{a}$}
\usepackage{tablefootnote}

\begin{document}

\title{Laser spectroscopy of aromatic molecules with optical cycling centers: strontium (I) phenoxides}
\author{Guanming Lao}
\affiliation{Department of Physics and Astronomy, University of California, Los Angeles, California 90095, USA}

\author{Guo-Zhu Zhu}
\email{guozhu.zhu@physics.ucla.edu}
\affiliation{Department of Physics and Astronomy, University of California, Los Angeles, California 90095, USA}

\author{Claire E. Dickerson}
\affiliation{Department of Chemistry and Biochemistry, University of California, Los Angeles, California 90095, USA}

\author{Benjamin L. Augenbraun}
\affiliation{Department of Physics, Harvard University, Cambridge, MA 02138, USA}
\affiliation{Harvard-MIT Center for Ultracold Atoms, Cambridge, MA 02138, USA}

\author{Anastassia N. Alexandrova}
\affiliation{Department of Chemistry and Biochemistry, University of California, Los Angeles, California 90095, USA}
\affiliation{Center for Quantum Science and Engineering, University of California, Los Angeles, California 90095, USA}

\author{Justin R. Caram}
\affiliation{Department of Chemistry and Biochemistry, University of California, Los Angeles, California 90095, USA}
\affiliation{Center for Quantum Science and Engineering, University of California, Los Angeles, California 90095, USA}

\author{Eric R. Hudson}
\affiliation{Department of Physics and Astronomy, University of California, Los Angeles, California 90095, USA}
\affiliation{Center for Quantum Science and Engineering, University of California, Los Angeles, California 90095, USA}
\affiliation{Challenge Institute for Quantum Computation, University of California, Los Angeles, California 90095, USA}

\author{Wesley C. Campbell}
\affiliation{Department of Physics and Astronomy, University of California, Los Angeles, California 90095, USA}
\affiliation{Center for Quantum Science and Engineering, University of California, Los Angeles, California 90095, USA}
\affiliation{Challenge Institute for Quantum Computation, University of California, Los Angeles, California 90095, USA}

\date{\today}

\begin{abstract}
We report the production and spectroscopic characterization of strontium (I) phenoxide ($\mathrm{SrOC}_6\mathrm{H}_5$, or SrOPh) and variants featuring electron-withdrawing groups designed to suppress vibrational excitation during spontaneous emission from the electronic excited state. Optical cycling closure of these species, which is the decoupling of vibrational state changes from spontaneous optical decay, is found by dispersed laser-induced fluorescence spectroscopy to be high, in accordance with theoretical predictions. A high-resolution, rotationally-resolved laser excitation spectrum is recorded for SrOPh, allowing the estimation of spectroscopic constants and identification of candidate optical cycling transitions for future work. The results confirm the promise of strontium phenoxides for laser cooling and quantum state detection at the single-molecule level.
\end{abstract}

\maketitle

\section*{Introduction }

Optical cycling transitions in atoms allow laser cooling of the center-of-mass motion, laser state preparation, and laser-induced fluorescence (LIF) state detection --- open-channel operations at the heart of many promising applications of quantum technology, including quantum computation~\cite{pino2021demonstration,debnath2016demonstration}, atomic clocks~\cite{ludlow2015optical,Brewer201927Alplus}, and quantum simulation~\cite{schafer2020tools,Monroe2021Programmable}. Optical cycling and cooling schemes have also been demonstrated in diatomic~\cite{DiRosa2004LaserCooling,Stuhl2008MagnetoOptical} and even some small polyatomic molecules~\cite{Isaev2016Polyatomic,kozyryev2016proposal}, including SrF~\cite{shuman2010laser}, YO~\cite{hummon20132d}, CaF~\cite{zhelyazkova2014laser,anderegg2017radio}, YbF~\cite{lim2018laser}, BaF~\cite{albrecht2020buffer, zhang2022doppler}, MgF~\cite{gu2022radiative}, AlF~\cite{hofsass2021optical}, SrOH~\cite{kozyryev2017sisyphus}, CaOH~\cite{baum20201d}, YbOH~\cite{augenbraun2020laser} and CaOCH$_3$~\cite{mitra2020direct}. Because they possess rich internal structures and complex interactions, molecules provide new opportunities in studies of dark matter detection~\cite{graham2011axion, van2015search}, measurement of electron's electric-dipole moment~\cite{hudson2011improved, doi:10.1126/science.1248213, acme2018improved}, parity violation tests~\cite{tokunaga2013probing, daussy1999limit}, and changes to fundamental constants~\cite{shelkovnikov2008stability, truppe2013search}.

The somewhat unexpected atom-like transitions supporting optical cycling and cooling in these small molecules have inspired searches for similar transitions in complex polyatomic molecules with an M-O-R structure \cite{kozyryev2016proposal}, where M is an alkaline-earth metal atom ionically bonded to oxygen (O) forming an optical cycling center (OCC) and R is a molecular ligand~\cite{ivanov2020two, klos2020prospects, augenbraun2020molecular}. In these molecules, the remaining metal-centered radical electron forms the highest-occupied
and the lowest-unoccupied molecular orbitals, HOMO and LUMO. For molecules with R having strong electron withdrawing capability, the HOMO and LUMO are localized on M, which typically indicates that the OCC is highly decoupled from the vibrational degrees of freedom. As a result, the diagonal vibrational branching ratio (VBR, which is to say the probability that spontaneous decay occurs on the 0-0 transition) is high, indicating that the spontaneous emission happens without a vibrational state change. This allows such molecules to repeatedly scatter photons before being pumped to the vibrational dark states, furnishing mechanical control and state detection of single molecules via laser illumination. 

Since optical cycling in this motif is predicted to be enhanced by the electron-withdrawing strength of the ligand, the diagonal VBR of M-O-R molecules could be tuned by functionalizing the ligand to promote this effect~\cite{Dickerson2021FranckCondon}. For example, according to a recent measurement of the VBRs~\cite{zhu2022functionalizing}, laser cooling of CaOPh-3,4,5-F$_3$ appears feasible from the perspective that each molecule could scatter $\approx$ 1000 photons with six to eight lasers. Compared to CaOPh, the three substitutions of $\mathrm{H} \!\rightarrow \!\mathrm{F}$ in the 3,4 and 5 positions on the ring enhances the electron-withdrawing strength of the ligand, rendering the Ca atom more ionic and thus suppressing spontaneous decays to excited vibrational states of the electronic ground state.

As molecules of M-O-R type, the strontinum variants, SrOPh-X, were also predicted to have high and tunable diagonal VBRs~\cite{Dickerson2021FranckCondon}. Compared to CaOPh-X, although the diagonal VBRs were predicted to be lower, the predicted difference is of the same order as the variation in measured VBRs of various calcium species \cite{zhu2022functionalizing}, suggesting that some of the strontium species may show better cycle closure if the variation is due to M-specific features. Further, Sr-containing molecules allow exploration of the role of stronger spin-orbit coupling \cite{kolkowitz2017spin} and nuclear spin structures \cite{zhang2014spectroscopic}. For the strontium variants, the excitation and repumping wavelengths can be directly produced by diode lasers.

Here, we report the production and spectroscopic characterization of strontium (I) phenoxide (SrOPh) and its derivatives, SrOPh-X (X = 3-CH$_3$, 3-F, 3-CF$_3$ and 3,4,5-F$_3$, see Scheme \ref{scheme}). Gas-phase molecules are produced by the reaction of Sr atoms generated by the ablation of Sr metal with the corresponding organic precursor vapor and cooled via collisions with the neon buffer gas in a cryogenic cell at a temperature of $\approx$ 23 K~\cite{sitext}. The first two electronically excited states, which have been proposed for optical cycling and laser cooling, are identified and the respective vibrational decays are observed using the dispersed laser-induced fluorescence (DLIF) spectroscopy. The diagonal vibrational branching ratios are estimated to be $0.82-0.96$, which indicates promise for laser cooling with a handful of vibrational repump lasers. To further characterize candidate optical cycling transitions, we have measured the rotationally-resolved excitation spectrum for the $\widetilde B - \widetilde X$ transition of SrOPh and obtained the molecular constants by fitting using \pgopher \cite{western2017pgopher}. 

\begin{scheme}
    \centering
    \includegraphics{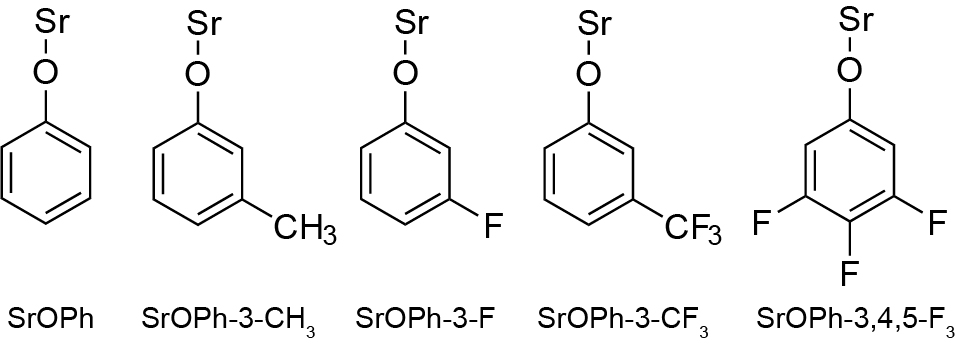}
    \caption{ Molecular structures of strontium (I) phenoxide and its derivatives studied in this work.
    }
   \label{scheme}
\end{scheme}

\begin{figure}
    \centering
    \includegraphics{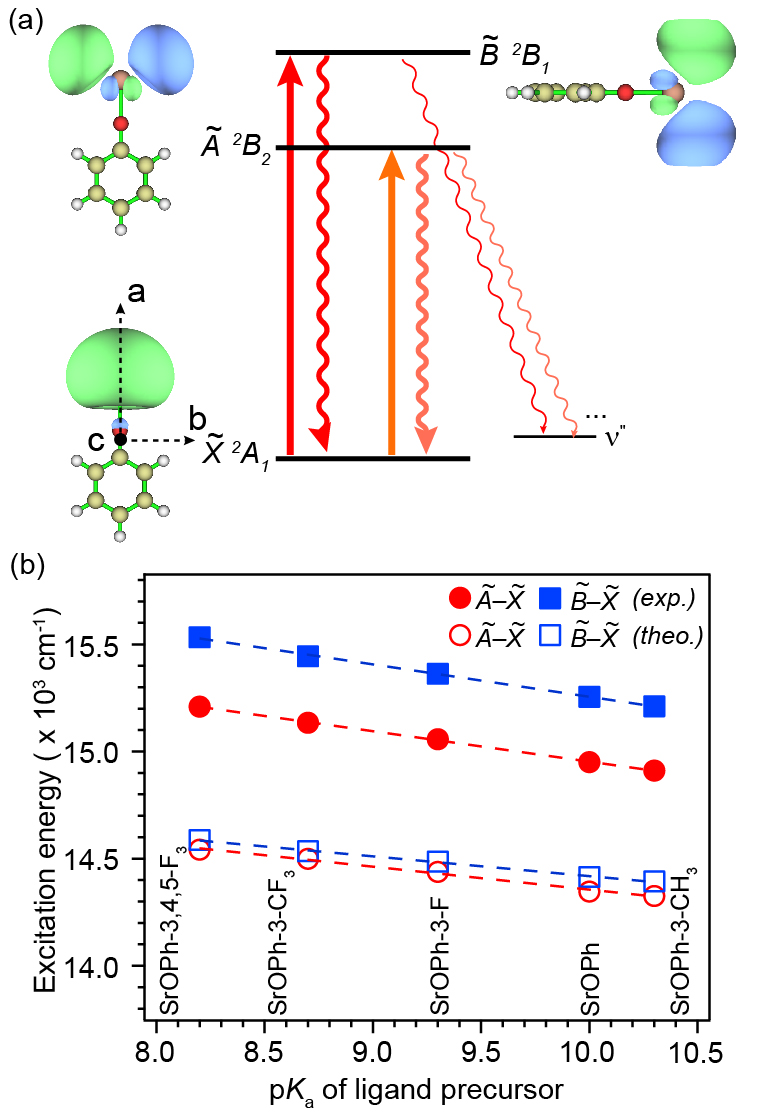}
    \caption{(a) Schematic energy levels of the transitions proposed for laser cooling. The molecular orbital and the respective symmetry of each state are illustrated for SrOPh with a C$_{2v}$ point group. For molecules with C$_s$ symmetry, the symmetries would be A$^\prime$ ($\widetilde X$), A$^{\prime\prime}$ ($\widetilde A$) and A$^\prime$($\widetilde B$). The principle inertial axes are also given. (b) Excitation energy versus \pKa\ for $\widetilde A - \widetilde X$ and $\widetilde B - \widetilde X$ transition for all studied species in an increasing order of ligand \pKa.  The linear fits of the experimental values yield $E_{\widetilde A - \widetilde X} = (16,372 - 142\,\text{\pKa})$ \cm and $E_{\widetilde B - \widetilde X} = (16,769 - 151\, \text{\pKa})$ \cm. }
    \label{fig:ex-pka}
\end{figure}

\section*{Results and discussion} 
In the calcium- and strontium- phenoxides, transitions to the two lowest electronic states ($\widetilde A$ and $\widetilde B$, Figure \ref{fig:ex-pka}a) have been proposed for laser cooling, since almost all photon scatters go back to the vibrationless ground state $\widetilde X$~\cite{Dickerson2021FranckCondon, zhu2022functionalizing,mitra2022pathway}. 
Figure \ref{fig:ex-pka}b ahows the measured transition energies of all molecules show a linear correlation with the acid dissociation constants, \pKa, of the precursor phenol. This linear trend has recently also been observed for CaOPh-X molecules \cite{zhu2022functionalizing, mitra2022pathway}. Lower \pKa\ implies higher electron-withdrawing capability of the R-O$^-$ ligand, which pulls the single electron away from the Sr atom, making it more ionic and increasing the HOMO-LUMO gap \cite{Dickerson2021FranckCondon}. Also shown are excitation energies calculated by time-dependent density functional theory \cite{sitext} which give a similar trend but systematically undershoot the excitation energies likely due to self-interaction error and approximate treatment of electronic correlation~\cite{acharya2018can}. The calculated energy gap of $\widetilde A - \widetilde B$  ($36-68$ \cm) is much smaller than the measured gap ($300-324$ \cm), similar to what was observed in CaOPh-X species but with a wider difference between the theory and measurement \cite{zhu2022functionalizing}. The theory-experiment discrepancies of the $\widetilde A - \widetilde B$ energy gap are likely due to the lack of spin-orbit coupling (SOC) in calculations \cite{liu2018rotational} and the wider difference in SrOPh-X is due to a stronger SOC effect in Sr.

\begin{figure}
    \centering
    \includegraphics{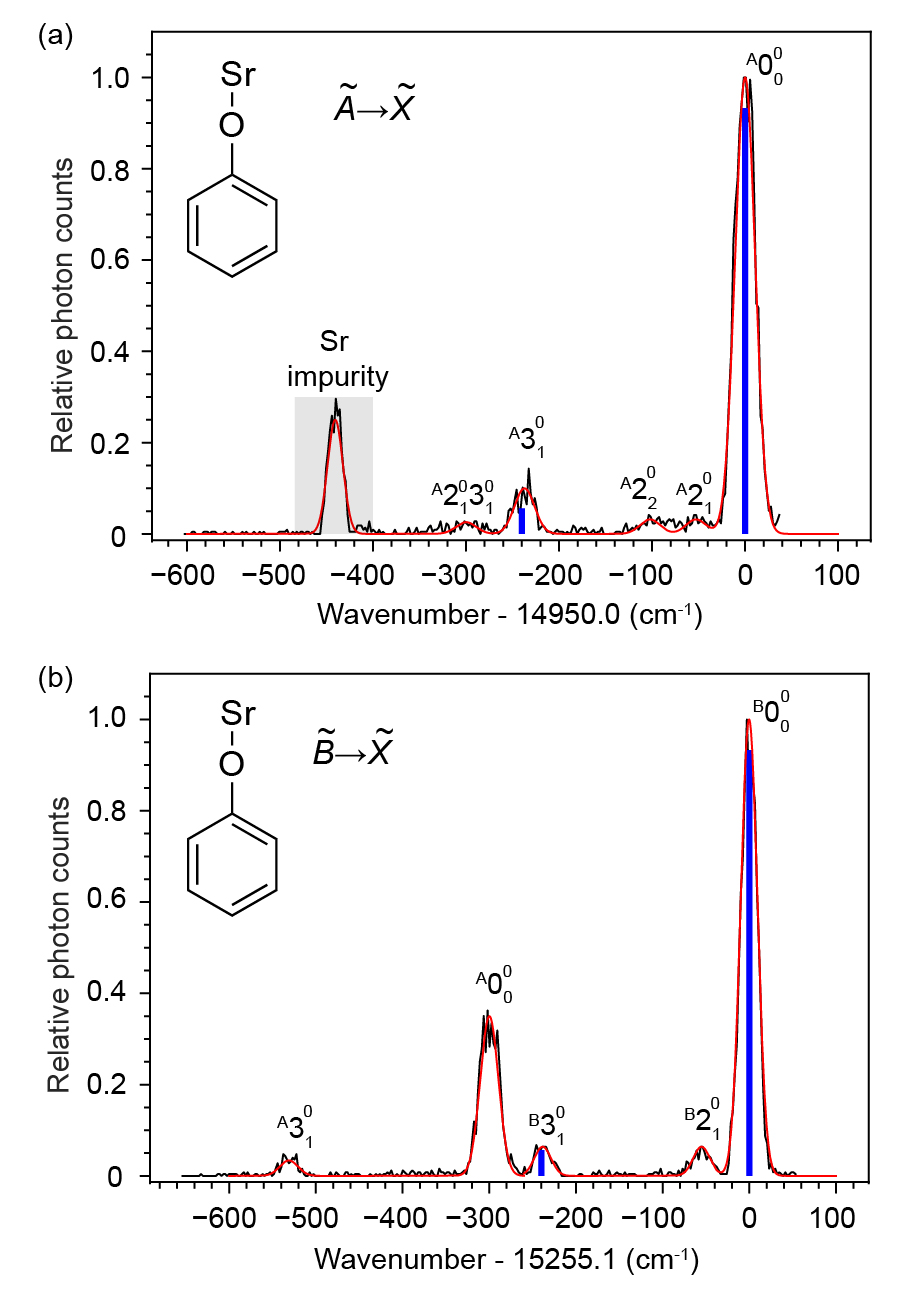}
    \caption{(a) and (b) Dispersed spectra of $\widetilde A \rightarrow \widetilde X$ and $\widetilde B \rightarrow \widetilde X$, respectively, of SrOPh excited by pulsed dye laser and measured by a spectrometer coupled with PMT. The experimental curves (black) are fitted with the Gaussian functions (red). The positions of the blue, vertical lines illustrate the theoretical frequencies while the intensities show the vibrational branching ratios of different vibrational modes of SrOPh. The Sr impurity peak in (a) is from the Sr emission of $5s5p~{}^3\mathrm{P}_1^o \rightarrow 5s^2~{}^1\mathrm{S}_0$ at 689 nm~\cite{nist}. The assignments of all resolved vibrational peaks are indicated.} 
    \label{fig:sroph-dlif}
\end{figure}

To measure the VBRs from the two electronic states, we performed DLIF spectroscopy of all molecules. Electronic excitation is provided by a pulsed dye laser (PDL) tuned to the 0-0 line and the spectrometer grating was scanned in time (over repeated excitation) to select the wavelength of LIF photons sent to a photomultiplier tube (PMT)~\cite{sitext}. Figure \ref{fig:sroph-dlif} shows the measured DLIF spectra of SrOPh while those of other species are presented in Figure \ref{fig:others-dlif}. Figure \ref{fig:sroph-dlif}a shows the spectrum of $\widetilde A~^2B_2 \rightarrow \widetilde X~^2A_1$ of SrOPh (Figure \ref{fig:ex-pka}a) at an excitation of 668.90 nm. The strongest peak at the origin, labeled as $^A0_0^0$, is due to the diagonal decay from $\widetilde A (v' = 0)$ to $\widetilde X (v'' = 0)$. The strong peak at $-440$ \cm~ is from excited atomic Sr created during laser ablation \cite{nist}. The peak at $-238$ \cm~is assigned to the strongest off-diagonal stretching mode $\nu_3$~(theo. 241 \cm) and the weak peak at $-54$ \cm~is assigned to the low-frequency bending mode $\nu_1$ (theo. 54 \cm). The other two weak peaks at $-100$ \cm~and $-297$ \cm, which do not match the calculated frequencies of any fundamental vibrational modes, are assigned to the overtone of the bending mode $^A2_2^0$ and a combinational mode of $^A2_1^03_1^0$, respectively. 

Figure \ref{fig:sroph-dlif}b shows the spectrum of $\widetilde B~^2B_1 \rightarrow \widetilde X~^2A_1$ of SrOPh (Figure \ref{fig:ex-pka}a) at 655.52 nm. Aside from the strongest diagonal peak $^B0_0^0$, four peaks are observed. The strong peak with a shift of $-300$ \cm~is due to a diagonal decay $^A0_0^0$ from the $\widetilde A$ state. The origin of the appearance of $^A0_0^0$ when exciting the $\widetilde B \leftarrow \widetilde X$ is unknown, but could be due to the collisional relaxation from $\widetilde B$ to $\widetilde A$ followed by fluorescence decay to the ground state $\widetilde X$ \cite{zhu2022functionalizing,mitra2022pathway,Liu2019collision}.  The identification of this feature as originating from the $\widetilde A$ state is further confirmed by the observation of the decay to the stretching mode $\nu_3$ at $-534$ \cm~from $\widetilde A$. The other two weak peaks, $-238$ \cm~ and $-55$ \cm, are due to the vibrational decay to the stretching mode $\nu_3$ and bending mode $\nu_1$, respectively. The full width at half maximum of all peaks is $\approx$ 22~ \cm~mainly due to the spectrometer resolution of approximately 20 \cm. Another measurement using a narrow-band continuous-wave (cw) laser to excite the $\widetilde B \leftarrow \widetilde X$ of SrOPh and an electron-multiplying charge-coupled device (EMCCD) camera to capture the fluorescence photons dispersed by the spectrometer.  This technique obtained a better spectral resolution ($\approx$ 5 \cm), allowing the resolution the combinational vibrational mode of $^B2_1^03_1^0$ (Figure \ref{fig:sroph_dlif_cw}), which is overlapped with the diagonal decay $^A0_0^0$ from the $\widetilde A$ state and not observed in Figure \ref{fig:sroph-dlif}b. The experimental and theoretical vibrational frequencies of all resolved fundamental modes are summarized in Table \ref{table:vib.freq.}.

\begin{figure*}
    \centering
    \includegraphics{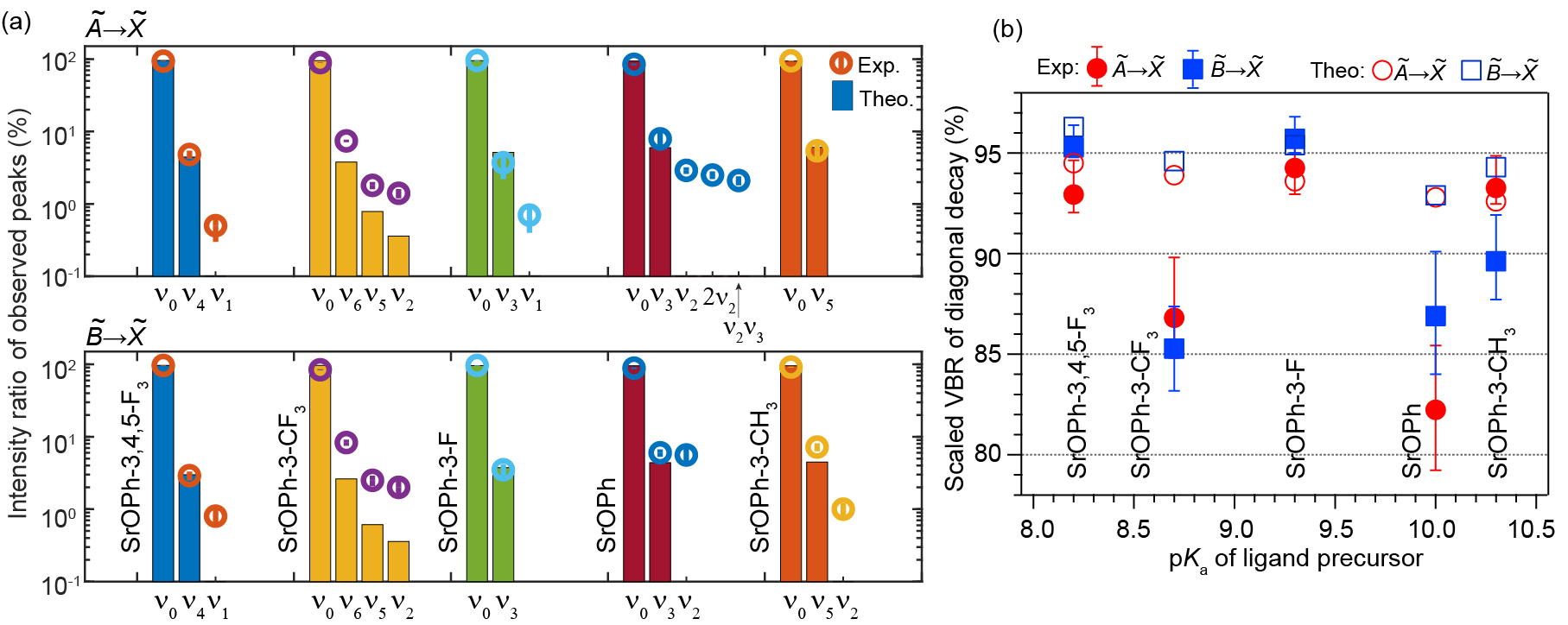}
    \caption{(a) Intensity ratio of observed decays for $\widetilde A \rightarrow \widetilde X$ and $\widetilde B \rightarrow \widetilde X$ transitions. Error bars are statistical errors from Gaussian fits. The vibrational label $\nu_i$ indicates the final vibrational modes of the $\widetilde X$ state. $\nu_0$ implys the decay that don't change vibrational state. (b) Scaled $0_0^0$ VBRs as a function of \pKa of all species. The scaling adds the contributions of those unobserved vibrational decays predicted by the theory to the observed intensity ratios of $0_0^0$ in (a). Error bars include the statistical errors from Gaussian fits and the systematic errors from the unobserved peaks~\cite{sitext}. } 
    \label{fig:vbr}
\end{figure*}

\begin{figure*}
    \centering
    \includegraphics{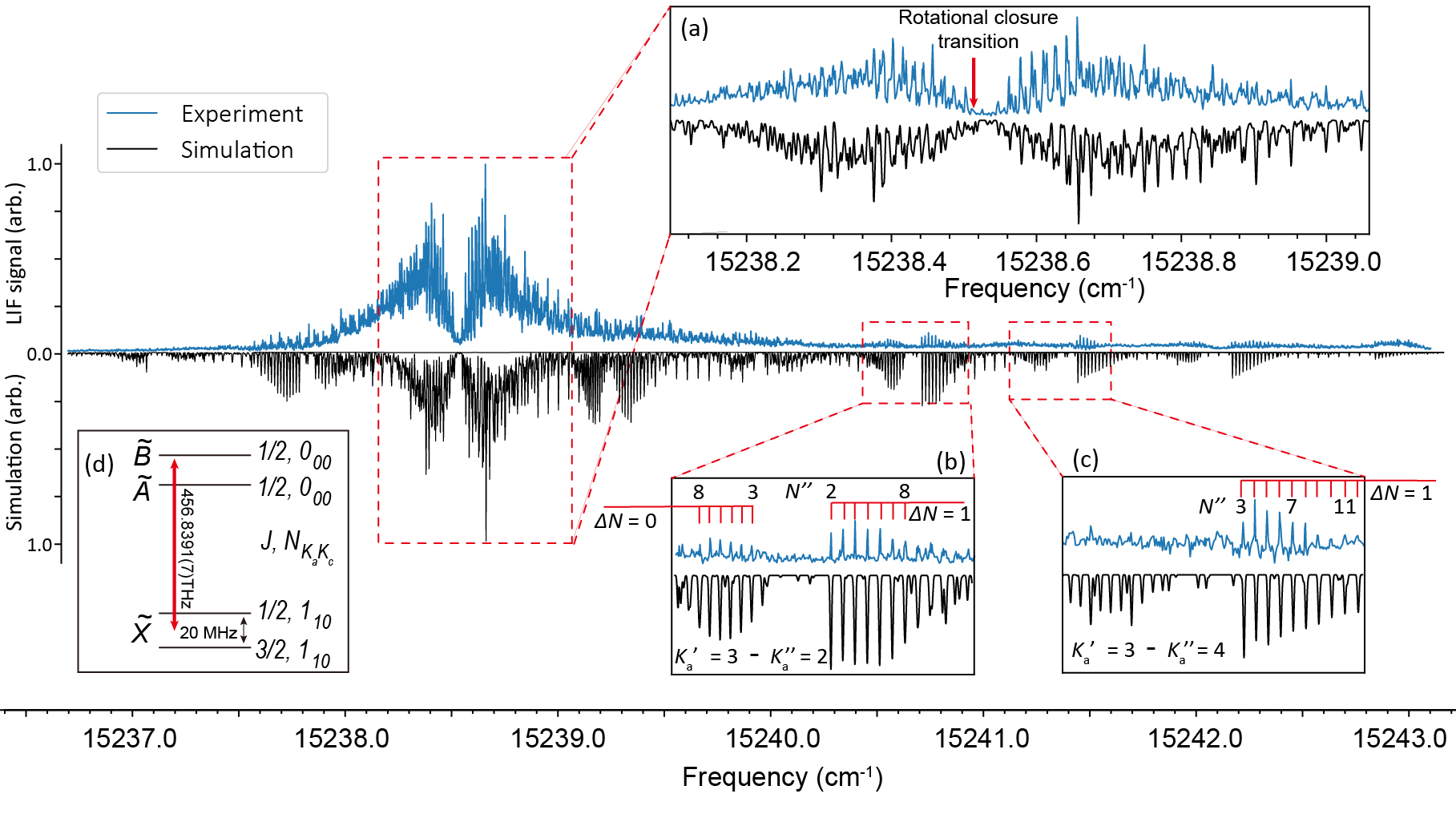}
    \caption{High-resolution rotationally-resolved excitation spectrum of the $\widetilde B \leftarrow \widetilde X$ transition of SrOPh. The upper  trace (blue) shows the experimental spectrum and the lower  trace (black) is the simulated spectrum with a Gaussian linewidth of 70 MHz and a rotational temperature $T_\mathrm{sim}=2.5$ K. Insets (a), (b) and (c) are expansions of some local features.  (a) displays detailed spectrum near 0-0 transition, while (b) and (c) show the $K_a'=3 \leftarrow K_a''=2$ and $K_a'=4 \leftarrow K_a''=3$ rotational bandheads, respectively. (d) shows the inferred position of the candidate rotational cycling transition between the spin-rotation manifold of the $N''=1$ state and $N'=0$ state.
    } 
    \label{fig:sroph-cw}
\end{figure*}

The relative heights of the peaks $^A0_0^0$ and $^B0_0^0$ in Figure \ref{fig:sroph-dlif} and Figures \ref{fig:others-dlif}-\ref{fig:sroph_dlif_cw} imply that both transitions are very diagonal with few vibration-changing decays. To extract the VBRs, all peaks are fitted with Gaussian functions, as shown by the red traces in Figure \ref{fig:sroph-dlif}, and the peak areas are extracted from the fits to obtain VBRs. A strict definition of VBR requires measurements of all vibrational decays. Due to finite measurement sensitivity ($\approx 10^{-2}$) and detection range ($< 600~$\cm), while we predict that our measurement is sensitive to the dominant leakage channels, the possibility of undetected decays contributes a systematic uncertainty on the measured VBRs.

For the vibrational decays that were identified for each molecule, and the ratios of line intensities to the total intensities of all \emph{observed} peaks are presented in Figure \ref{fig:vbr}a. In both electronic transitions, the relative ratios of observed peaks show good agreement with the calculated VBRs. The vibrational decays to the strongest off-diagonal Sr-O stretching mode ($\nu_3$, $\nu_4$, $\nu_5$ or $\nu_6$) and the low-frequency bending mode ($\nu_1$ or $\nu_2$) have been observed for all molecules. The theoretical VBRs of the low-frequency bending modes are underestimated possibly due to the vibronic coupling and anharmonicity effect not considered in the calculation~\cite{zhu2022functionalizing, Dickerson2021FranckCondon}. SrOPh also shows unpredicted decays to the overtone of mode $\nu_2$ and a combinational mode $\nu_2\nu_3$ where the intensities could be from the vibronic coupling. The intensity ratios of all observed decays are summarized in Table \ref{table:modes-vbr}. Figure \ref{fig:vbr}b plots the estimated VBRs of the diagonal peak $0_0^0$ of each transition as a function of \pKa. The scaled $0_0^0$ VBRs are obtained by adding the estimated contribution of the unobserved peaks predicted by the theory to the normalized intensity ratios of the observed $0_0^0$ intensities~\cite{sitext}. Both SrOPh-3-F and SrOPh-3,4,5-F$_3$ molecules show VBRs $>$ 95\% for the $\widetilde B \rightarrow \widetilde X$ transition and $>$ 90\% for the $\widetilde A \rightarrow \widetilde X$ transition, while SrOPh has the lowest VBR of 82.2\% for $\widetilde A \rightarrow \widetilde X$ transition.

The VBRs for SrOPh-3-CF$_3$ shows the largest discrepancy between the calculation and the measurement, potentially due to the larger vibronic mixing between the $\widetilde A$ and $\widetilde B$ caused by the low symmetry and large electron inductive effect from the CF$_3$ group~\cite{zhu2022functionalizing}. The error bars include both the statistical uncertainties from the Gaussian fit and the systematic uncertainty estimate from the unobserved peaks. Three additional systematic errors, including signal drift during measurement, the wavelength response of the spectrometer, and the diagonal excitation from the vibrationally excited states, are estimated to be a few percent in total~\cite{sitext}. 

To further investigate the potential of these species for optical cycling, a high-resolution excitation spectrum (obtained by collecting LIF as a continuous-wave (cw) excitation laser is scanned) of SrOPh for the $\widetilde B (v' = 0) \leftarrow \widetilde X (v'' = 0)$ transition is recorded at a step size of $25 - 50$ MHz in a cryogenic buffer-gas beam (CBGB) \cite{hutzler2012buffer,sitext} and fitted with \pgopher \cite{western2017pgopher}, as presented in Figure \ref{fig:sroph-cw}. Since SrOPh is an asymmetric-top molecule, the rotational states are labeled as $N_{K_aK_c}$, where $N$ is the rotational angular momentum, $a$ and $c$ label the inertial axes lying along the Sr-O bond and perpendicular to the molecular plane (Figure \ref{fig:ex-pka}a), respectively, $K_a$ and $K_c$ are the projection of $N$ onto the two axes in the prolate and oblate limits, respectively. Figure \ref{fig:sroph-cw}a shows the expansion of the two congested bands at 15238.5 \cm, while Figures \ref{fig:sroph-cw}b,c show two well-resolved rotational bands. A full rotational analysis is difficult due to the high density of rotational lines in the middle of the spectrum (Figure \ref{fig:sroph-cw}a) , but the individually resolved lines (Figures \ref{fig:sroph-cw}b,c) make it possible to fit the spectrum to extract some spectroscopic constants.

\begin{table}[t!]
    \centering
    \begin{tabular}{c | c c | c c}
    \hline
    \hline
         \multirow{2}{*}{Constant} & \multicolumn{2}{c}{$\widetilde B~^2B_2$  } &\multicolumn{2}{|c}{$\widetilde X~^2A_1$} \\ \cline{2-5} & Exp. & Cal.  & Exp.& Cal.\\ \hline
         
         $T_0$ & 15238.7155(23) & ~ & ~ & ~ \\ 
        $A$ & 0.1923(6) & 0.1915   & 0.1934(11) & 0.1916 \\ 
        $\frac{1}{2}(B+C)$ & 0.01520(36) & 0.01522  & 0.01508(36) & 0.01513 \\ 
        $(B-C)\times 10^3$ & 1.28(20) & 1.21  & 1.13(12) & 1.19 \\ 
        $\epsilon_{aa}$ & $-0.6894(6)$ &  & - & ~ \\ 
        $\epsilon_{bb}\times 10^3$ & 34(10) & ~  & 1.3(1.7) & ~ \\ 
        $\epsilon_{cc}\times 10^3$ & 16(7) & ~  & $-1.3(1.8)$ & ~ \\ 
        $D_N \times 10^8$ & $-14(8)$ & ~  & $-14(8)$ & ~ \\ 
        $D_{NK} \times 10^7$ & $-5(11)$ & ~  & $-28(23)$ & ~ \\ 
        $D_K\times 10^4$ & 1.3(5)  & ~  & 5.2(1.1) & ~ \\ 
        $H_K \times 10^6$ & 3.0(1.4) & ~ & 21(4) & ~ \\ 
        
        \hline
    \end{tabular}
    \begin{tablenotes}
    \item $T_0$: electronic transition energy; $A, B,$ $C$: molecular rotational constants; $\epsilon_{aa}, \epsilon_{bb},\epsilon_{cc}$: spin-rotation coupling constants; $D_{N},$ $D_{NK},$ $D_{K}$: centrifugal distortion constants; $H_K$: sextic centrifugal distortion correction.
    \end{tablenotes}
    \caption{Molecular constants of SrOPh obtained by fitting the rotationally-resolved excitation spectrum in Figure \ref{fig:sroph-cw} with \pgopher. All quantities are presented in \cm.  
    }
    \label{table:mol.const}
\end{table}

Using a custom program to fit the spectral contour and \pgopher \cite{western2017pgopher} to refine and iterate the line assignments~\cite{sitext}, we have assigned nearly 400 rotational transitions and obtained the final fitted spectrum given as the black traces in Figure \ref{fig:sroph-cw}. The fitting is in agreement with the experimental measurement for the middle broad bands and the $K_a'=3 \leftarrow K_a''=2$ and $K_a'=4 \leftarrow K_a''=3$ bands, as expanded in Figures \ref{fig:sroph-cw}a-c. The best fit molecular constants, including the transition energy, rotational constants, spin-rotational constants and centrifugal distortion corrections, are reported in Table \ref{table:mol.const}. The measured rotational constants are in good agreement with the calculated values. The spin-rotation constant $\epsilon_{aa}$ in the ground state is too small to be determined from the spectrum, and $\epsilon_{aa}$ in the $\widetilde B$ state is large because of the coupling to the $\widetilde A$ state. The larger value of spin-rotational constant than the rotational constants in $\widetilde B$ implys a strong SOC effect apart from the direct coupling between the spin and molecular rotation. Based on the second order perturbation theory \cite{van1951coupling, morbi1997high} and the measured constants, the SOC constant in SrOPh is estimated to be $\approx 272$ \cm, which is close to that of SrOH ($A^2\Pi$, $\approx 265$ \cm)~\cite{PRESUNKA199597}. The large SOC also dominates the energy separation of $\widetilde A - \widetilde B$, elucidating the discrepancy between the calculation and the measurement in Figure \ref{fig:ex-pka}b~\cite{sitext,liu2018rotational}.

While involving more parameters has been able to enhance the accuracy of fitting, many parameters in such scenarios tended to fit to values consistent with zero, and we therefore omit those in our analysis. The rotational temperature from the fit is 2.5 K \cite{sitext}. The colder temperature is due to the free expansion of neon buffer gas from the cryogenic cell ($\approx 23$ K) to form a beam with SrOPh entrained~\cite{hutzler2012buffer}. As the SrOPh $\widetilde B \leftarrow \widetilde X$ transition dipole moment lies along the principle axis \textit{c} (Figure \ref{fig:ex-pka}b), the rotationally closed photon cycling transition is the \textit{c}-type transition $N'_{K_a'K_c}=0_{00} - N''_{K_a''K_c''} = 1_{10} $~\cite{augenbraun2020molecular}, which is estimated to be at $456.8391(7)$ THz based on the fitting results and shown in Figure \ref{fig:sroph-cw}d.

\section*{Conclusion} 
In summary, we have produced strontium (I) phenoxide (SrOPh) and derivatives featuring electron-withdrawing groups in a cryogenic cell. Two proposed laser cooling transitions ($\widetilde A - \widetilde X$ and $\widetilde B - \widetilde X$) of each molecule have been identified and the transition energies show linear trends as the ligand \pKa, which can be used to look for  transitions of new molecules containing Sr. The overall vibrational branching ratios considering contributions of unobserved vibrational decays are estimated to be $82.2\% - 95.8\%$. Among them, SrOPh-3-F and SrOPh-3,4,5-F$_3$ molecules show diagonal VBRs $>$ 95\%, potentially enabling laser cooling with fewer than ten vibrational repumping lasers. The rotationally-resolved spectrum for the $\widetilde B \leftarrow \widetilde X$ transition of SrOPh is presented and molecular constants are obtained. The spin-orbit interaction that couples the $\widetilde A$ and $\widetilde B$ states is estimated to be 275 \cm, which has a strong effect on the energy splitting of $\widetilde A - \widetilde B$. The rotational closure transition for optical cycling is estimated to be centered near 456.8391 (7) THz.  This work paves the way for optical cycling of SrOPh and other large molecules using diode lasers.

\emph{Acknowledgements --} The authors thank John Doyle and Timothy Steimle for helpful discussions. This work was supported by the AFOSR (grant no. FA9550-20-1-0323), the NSF (grant no. OMA-2016245, PHY-2207985 and DGE-2034835), NSF Center for Chemical Innovation Phase I (grant no. CHE-2221453). This research is funded in part by the Gordon and Betty Moore Foundation.  Computational resources were provided by XSEDE and UCLA IDRE shared cluster hoffman2.

\bibliographystyle{apsrev4-1_no_Arxiv}
\bibliography{SrOPh_main}

\section*{Supplementary information}

\subsection{Experimental methods}
\textbf{Molecule production.} All SrOPh-X molecules were produced by the reaction of Sr atoms generated by laser ablation of a metallic Sr metal chunk using the Minilite pulsed Nd:YAG laser at 1064 nm (pulse energy $\approx 6$ mJ, repetition rate 10 Hz) with different ligand precursors in a cryogenic buffer-gas cell operated at $\approx$~23~K. Five ligands -- phenol, $m$-cresol, 3-fluorophenol,  3-(trifluoromethyl)phenol and 3,4,5-trifluorophenol -- purchased from Sigma Aldrich were individually heated in a reservoir to supply the respective vapors, which were guided via a heated gas line into the cryogenic cell with a density $\approx 10^{13}$~cm$^{-3}$. The reaction products were then cooled by colliding with a neon buffer gas of density $\approx 10^{15-16}$~cm$^{-3}$.

\noindent\textbf{DLIF measurement in the cryogenic cell.} The resulting SrOPh-X molecules have two low-lying electronic states proposed for laser cooling. To look for those states, a tunable, pulsed dye laser (10 Hz, LiopStar-E dye laser, linewidth 0.04~cm$^{-1}$ at 620~nm) were used to excite molecules in the cryogenic cell and the laser wavelength was scanned. When the laser wavelength hit electronic resonance, the molecule was excited and followed by the emission of molecular fluorescence. The fluorescence was then collected via an imaging system into a model 2035 McPherson monochromator equipped with a 1200 lines/mm grating and detected by a PMT. The dispersed measurement were done by parking the laser wavelength at the electronic resonance and continuously scanning the grating of the spectrometer at an increment of 0.10 nm while monitoring the fluorescence photons. The entrance and exit slit widths were both set at 0.20 mm, resulting in a spectrometer resolution of $\approx 20$ \cm.

A different method was also used to measure the DLIF spectrum for SrOPh $\widetilde B \rightarrow \widetilde X$ transition (Figure \ref{fig:sroph_dlif_cw}). In this method, the SrOPh molecules were excited by a cw laser illuminated from a home-build external-cavity diode laser (ECDL). The fluorescence was dispersed by the monochromator at a fixed grating position and detected by an EMCCD camera. The entrance slit of the spectrometer was set to 0.03 mm to achieve a resolution of $\approx$ 3~\cm. In this measurement, all fluorescence photons were collected simultaneously by the EMCCD, which avoided the systematic errors due to the ablation or PDL energy drift in scanning the grating in the first method.  

\noindent\textbf{High resolution excitation spectroscopy of SrOPh $\widetilde B - \widetilde X$.} To look for the rotational closure transitions, We performed the high-resolution excitation spectroscopy of SrOPh in a molecular beam. SrOPh molecules formed in the cryogenic cell were extracted out via a 9 mm cell aperture and entrained into a neon buffer gas beam. The excitation zone is $\approx$ 23 cm downstream the cell aperture. A cw laser from an ECDL was scanned with an increment of $25-50$ MHz near the $\widetilde B (v' = 0) - \widetilde X (v'' = 0)$ transition of SrOPh determined by the low-resolution PDL measurement. The fluorescence were collected by a PMT at the perpendicular direction. Due to the cooling effect in the expansion, SrOPh molecules in the beam are colder than those thermalized in the cell at a temperature $\approx$ 23 K.

\subsection{Theoretical methods}
Molecular geometries, excitation energies, and Franck-Condon factor (FCFs) calculations were performed in Gaussian16~\cite{frisch2016gaussian}.  Density functional theory (DFT) was used for the ground states while time-dependent DFT was used for excitation energies of excited states.  A superfine grid was used with the PBE0-D3 functional with dispersion corrections and the def2-TZVPPD basis set~\cite{perdew1996rationale,weigend2005balanced,grimme2010consistent,rappoport2010property}. An effective core potential (ECP) was used for the Sr atom within the def2-TZVPPD basis set. Molecular orbitals were generated with an isosurface of 0.03 in the Multiwfn program~\cite{lu2012multiwfn}. The FCFs calculated included Duschinsky rotations, which seems to be sufficient for an overall trend in FCF. However, the anharmonicity and vibronic coupling effects do play a key role in the the FCFs of the low-frequency bending modes \cite{Dickerson2021FranckCondon,zhu2022functionalizing}, which were underestimated here in comparison to the experimental measurements.  
The calculated FCFs can be converted to the VBRs using the formula \cite{augenbraun2020molecular,zhu2022functionalizing}:
\begin{align}
    b_{iv',fv''} & = \frac{A_{iv',fv''}}{\sum_{fv''}A_{iv',fv''}}\nonumber\\
    & = \frac{|\mu_{iv',fv''}|^2~\times~(\nu_{iv',fv''})^3}{\sum_{fv''} |\mu_{iv',fv''}|^2~\times~(\nu_{iv',fv''})^3}\nonumber\\
    &\approx \frac{\text{FCF}_{iv',fv''}~\times~ \nu^3_{iv',fv''} }{\sum_{fv''} \text{FCF}_{iv',fv''} ~\times~\nu^3_{iv',fv''}}
\end{align}

\noindent where $i$ and $f$ imply the initial and final states, respectively. $b_{iv',fv''}$ is the branching ratio, $A_{iv',fv''}$ is the Einstein coefficient for spontaneous emission, $\mu_{iv',fv''}$ is the transition dipole moment and $\nu_{iv',fv''}$ is the transition frequency.  

\subsection{DLIF spectra of other molecules}
Figure \ref{fig:others-dlif} presents the DLIF spectra of other four molecules, which were recorded by monitoring the fluorescence signal from the respective excited states when scanning the grating wavelength at a step size of 0.1 nm. All peaks were fitted with Gaussian functions, as shown by the red traces overlapped with the experimental black traces. Comparing to the theoretical vibrational frequencies and the respective VBRs (blue lines), those peaks can be readily assigned. Take SrOPh-3-CH$_3$ for an example, two peaks show up in the $\widetilde A \rightarrow \widetilde X$ decay in Figure \ref{fig:others-dlif}a. The origin peak labeled as $^A0_0^0$ represents the decay of $\widetilde A (v' = 0) \rightarrow \widetilde X (v'' = 0)$, while a weak peak at a frequency shift of $-226$ \cm~matches well with the theoretical vibrational frequency of the Sr-O stretching  mode $\nu_5$ (230 \cm), which is also the most off-diagonal vibrational mode. In Figure \ref{fig:others-dlif}b of $\widetilde B \rightarrow \widetilde X$ decay, except for the diagonal peak ($^B0_0^0$) at the origin and the stretching-mode peak ($^B5_1^0$) at $-230$ \cm, two additional peaks are observed. The strong peak at $-300$ \cm~is assigned to $^A0_0^0$, which is always observed and due to the collisional relaxation from $\widetilde B$ to $\widetilde A$ followed by the radiative decay. A very weak peak at $-43$ \cm~is the decay to the low-frequency bending mode $\nu_2$ (50 \cm). The similar spectra features of decays to the most-off diagonal Sr-O stretching mode and the low-frequency bending mode and the appearance of the $^A0_0^0$ peak when exciting to $\widetilde B$ have been observed for all other molecules. A more complex decay scenario is observed for SrOPh-3-CF$_3$ mainly due to the low-symmetric structure introduced by the electron-withdrawing group of CF$_3$. More vibrational decays have been observed. In Figure \ref{fig:others-dlif}e, three fundamental vibrational modes have been resolved from the decay of $\widetilde A$ and assigned to $\nu_2$, $\nu_5$ and $\nu_6$. The decay from the $\widetilde B$ state is more complicated due to the relaxation decay to the  $\widetilde A$ state. Except for the same vibrational decays from $\widetilde B$, three more peaks are observed. The strongest peak at  $-300$ \cm~is due to the collisional relaxation and fluorescence pathways of $\widetilde B \rightarrow \widetilde A \rightarrow \widetilde X$. The high intensity is caused by the large CF$_3$ group which increases the collisional relaxation rate from the $\widetilde B$ state. Two other weak peaks are assigned to $^A5_0^1$ $^A6_0^1$ due to the relaxation from $\widetilde B$ to the vibrationally excited states of $\widetilde A$ followed by the fluorescence decays. The vibrational frequencies and VBRs of all observed vibrational modes are given in Tables \ref{table:vib.freq.} and \ref{table:modes-vbr}.

\subsection{Error analysis of VBRs}
All observed peaks in DLIF spectra in Figures \ref{fig:sroph-dlif} and \ref{fig:others-dlif} are fitted with the Gaussian function using the parameters of peak location, height and width. For the $i$th peak in each spectrum, the peak area ($M_i$) is extracted and the error of the area ($\delta M_i$) is estimated from the covariance matrix of the fitting parameters. The intensity ratios of each peak, as shown in Figure \ref{fig:vbr}a, is obtained from the ratio of $M_i$ and the total area of all the observed peaks, $\sum_{i=0}^pM_i$. The statistical error of each intensity ratio is then calculated from the relative uncertainties $\delta M_i/M_i$. Besides the statistical fitting errors, several sources of systematic error in the DLIF measurement are discussed and listed in Table \ref{table:syserr}. 

The first systematic error come from the unobserved peaks which contribute to the VBRs. A true VBR depends on contributions of all possible decay pathways. Due to a low measurement sensity and a small detection window, only a few vibrational decays have been observed for each transition. Compared to a complete description of vibrational decays obtained from calculated FCFs, all unobserved vibrational decays are therefore a source of the systematic uncertainty, which is estimated by \cite{zhu2022functionalizing}: 
\begin{equation}
\begin{aligned}
  S'_{0}=\frac{S_{0}}{\sum_{\substack{i=0}}^{p}S_{i} + \sum_{\substack{i=p+1}}^{N}\frac{T_{i}}{C}},
\end{aligned}
\end{equation}

\noindent where $S'_0$ is the scaled diagonal VBR considering contributions from the unobserved vibrational decays, $S_0$ is the observed intensity ratio of the diagonal peak, $S_i$ is the observed intensity ratio of the $i$th vibational decay, $T_i$ is the calculated VBR of the $i$th vibational decay and $C$ is a scaling factor that averages the ratio of theoretical VBR ($T_i$) to experimental intensity ratio ($S_i$) for all observed peaks. Since the VBRs for bending modes are underestimated by theory, $T_0$ is usually larger than or roughly equal to $S_0$, thus the lowest scaled VBR $S''_0$ can be obtained when excluding the $T_0$/$S_0$ in the scaling factor and used as an estimate for the lower bound value of the diagonal VBR. For the measured intensity ratio $S_0$ is always overestimated and used as the upper bound of the diagonal VBR, while the scaled VBR $S'_0$ is used as the plot data points in Figure \ref{fig:vbr}. As summarized in Table \ref{table:syserr}, the uncertainties of the unobserved peaks as differences of $S_0$ and $S''_0$ are in the range of $1.1-3.2$\%.   

Another source of systematic uncertainty is the signal drifting in the measurement due to the change of experimental conditions. Except for the DLIF measurment of SrOPh $\widetilde B \rightarrow \widetilde X$ by the cw laser and EMCCD (Figure \ref{fig:sroph_dlif_cw}), all other DLIF spectra were taken by scanning the grating of the spectrometer to disperse photons onto a PMT. This means that the fluorescence photons at different wavelengths were not detected simultaneously. A typical scan of 15 nm wavelength range would take 75 minutes with an increment of 0.1 nm and 300 averages for each wavelength. During the scan, the signal was slowly drifting mainly due to the dust accumulation on the imaging lens and the pulse intensity drifting from both the ablation and the excitation lasers. We kept track of the signals before and after the whole scan and found that the signal change can vary by up to 20\%, which can lead to an error of 1.0\% in VBR estimation as such signal change mainly affects the off-diagonal vibrational transition signal. In addition, the dispersed photons at different wavelength were detected simultaneously in the EMCCD measurement, which eliminates the error due to signal drifting. The error can be estimated by the difference of the diagonal VBR of SrOPh $\widetilde B \rightarrow \widetilde X$ from the two different methods, which is 1.2 \%.

As discussed in the error analysis of CaOPh-X \cite{zhu2022functionalizing}, the wavelength response of the spectrometer and the imperfection of the mirrors and lenses in the imaging system could cause a systematic error up to $\approx$ 1\%. The last error source comes from the diagonal excitation of vibrationally excited modes in the ground state. Due to a cell temperature of $\approx$ 23 K, the thermalized molecules can have thermal populations of $\approx 5\%$ of the low-frequency bending mode and $\approx$ 10$^{-6}$ of the stretching mode. Those vibrationally excited states in the $\widetilde X$ state could be near-resonantly excited by the pulsed dye laser to the same vibrational levels of the upper states. The following decays from those diagonal excitations can cause an error up to 0.5\%. 

By adding the four systematic errors in quadrature, a total systematic uncertainty is estimated to be $1.9\% - 3.5\%$. Considering the statistical uncertainties from the Gaussian fits, the final upper and lower bounds of uncertainties for the diagonal $0_0^0$ VBRs are obtained and plotted in Figure \ref{fig:vbr}b. 

\subsection{Fitting of high-resolution excitation spectrum of SrOPh}
The rotationally-resolved excitation spectrum of SrOPh has been fitted to estimate the molecular constants. The energy levels of SrOPh $\widetilde B$ and $\widetilde X$ states are computed from an effective Hamiltonian which includes rotation, electron spin-rotation coupling and centrifugal distortion correction terms:
\begin{equation}
    H_\mathrm{eff} = H_\mathrm{Rot}+H_\mathrm{SR}+H_\mathrm{cd}.
\end{equation}
The rotational Hamiltonian is
\begin{equation}
    H_\mathrm{Rot}=AN_a^2+BN_b^2+CN_c^2,
\end{equation}
the spin-rotation coupling term is
\begin{equation}
    H_\mathrm{SR} = \epsilon_{aa}N_aS_a+\epsilon_{bb}N_bS_b+\epsilon_{cc}N_cS_c,
\end{equation}
and the centrifugal distortion correction used in this work is
\begin{equation}
    H_\mathrm{cd} = -D_NN^2(N+1)^2-D_{NK}N(N+1)K_a^2-D_KK_a^4+H_KK_a^6,
\end{equation}
here the sextic centrifugal distortion correction $H_K$ is included to enhance the fitting quality of the bandheads of the observed $\Delta K_a = \pm1$ bands. 

Although a few rotational bands recognized in the SrOPh excitation spectrum can determine a few rotational constants in good accuracy, the high density of rotational line near 0-0 transition ($\sim 10^2$/GHz in the computing limit of $J_{max}=30$) is the main challenge for fitting the rest of the parameters. Therefore, we used a homemade program first to search for parameters that can roughly fit the spectrum in contour and limit the searching range of the parameters for \pgopher's subsequent fitting~\cite{western2017pgopher}. To avoid the program being trapped by local minima, two algorithms were used in turns: the mini-batch gradient descent (MBGD) \cite{konevcny2015mini} and the genetic algorithm (GA)~\cite{meerts2004new}. MBGD works similarly as the well-known gradient descent method, while its gradient is computed from a batch of randomly chosen data instead of the whole data set during each iteration to jump out of local minima with semi-stochastic steps. MBGD is the main algorithm that searches for a potential solution iteratively, and GA checks whether the MBGD result is optimal within a larger parameter space. The target functions of the two algorithms are different, with the purpose of making their local minima to be also different.

A contour fitting result is accepted as the initial input of \pgopher if it is agreed by both MBGD and GA. In \pgopher, the fitting of molecular constants is done in following procedure: first, the clearly observed bandheads are matched to the simulated rotational bands, such as the lines labeled in Figure \ref{fig:sroph-cw}b and \ref{fig:sroph-cw}c. With such assignment \pgopher can calculate parameters $T_0$, $A$, $\Bar{B}$, $\epsilon_{aa}$, $D_K$ and $H_K$ in a better accuracy than the contour fitting. Next, the rest of the parameters are obtained by some details of the spectrum near the 0-0 transition, such as the spacing of lines in different rotational bands, shape of peaks, and the order of line strength. For concreteness, some strong peaks in the middle of the spectrum (Figure \ref{SrOPh-cw_assignment}a) are assigned to the transitions with $K_a''=0, 1, 2$ and different P, Q and R branches (P, Q, and R labels refer transitions with $\Delta J = -1, 0$ and $1$, respectively): the strongest few peaks in Q branch are assigned to some observed peaks, see Figure \ref{SrOPh-cw_assignment}b; and for the P and R branch transitions, the line assignment could be made according to some local features, see Figure \ref{SrOPh-cw_assignment}c. 

Fine adjustment of fitting is achieved by adjusting the parameters, re-assigning some lines or bands tentatively, and running the fit based on the updated assignment. This procedure should be repeated multiple times before the parameters become converged.

The rotational temperature in \pgopher simulation is set based on the normalized strength of the rotational bandhead of $K'_a = 6 \leftarrow K''_a = 5$, the farthest band we can recognize in experiment. It is found to be close to the experimental results when $T_\mathrm{sim}$ is around $2-3$ K. In the contour fitting procedure, the rotational temperature is estimated to be $4-7$ K, depending on the fitting condition such as linewidth and the upper limit of the rotational quantum number $J_{max}$. 

The estimated molecular constants reported in Table \ref{table:mol.const} are from the best fit result whose simulated spectrum pattern matches most of observed peaks near the 0-0 transition, with the assignment error bars calculated from the standard errors of the estimated values given by multiple fitting attempts. These attempts follow the same rotational bandhead assignment in the first step and have a similar P, Q, R line distributions depicted in Figure \ref{SrOPh-cw_assignment}, while the numbers and the positions of assigned lines near the 0-0 transition are varied to reflect the parameter fluctuations caused by different assignments. 

\subsection{Spin-orbit coupling effect in SrOPh} 
According to the second order perturbation theory \cite{van1951coupling, morbi1997high}, the relation between the SOC constant $A_\mathrm{so}$ and the effective spin-rotation constant of the $\widetilde{B}$ state could be estimated by the following equation:
\begin{equation}
\label{eaaAso}
    \epsilon_{aa}^{\widetilde{B}} \approx -\frac{4AA_\mathrm{so}}{E_{\widetilde{B}}-E_{\widetilde{A}}}.
\end{equation}
With the measured constants, the SOC constant in SrOPh estimated by Eq.(\ref{eaaAso}) is $A_\mathrm{so}\approx 272$cm$^{-1}$, which is similar to the $A_\mathrm{so}$ of the SrOH $A^2\Pi$ state  ($\approx 265$ cm$^{-1}$) \cite{PRESUNKA199597}. 

The SO interaction can also explain why the energy gaps between $\widetilde{A}$ and $\widetilde{B}$ states ($\approx 300$ cm$^{-1}$) of all the species shown in Figure \ref{fig:ex-pka}b are much larger than their prediction. In the calculation, the electronic energies are computed without SOC effect, and the excited state wavefunctions, conventionally labeled as $|A^2B_2\rangle$ and $|B^2B_1\rangle$, are assumed to follow the $C_{2v}$ symmetry. However, the SO interaction strongly mixes the two states and breaks the $C_{2v}$ symmetry in these wavefunctions. We can use a simple two level system model to demonstrate such mixing, in which the total electronic Hamiltonian is contributed by the electronic Hamiltonian ($H_\mathrm{el} = E_A|A^2B_2\rangle\langle A^2B_2|+E_B|B^2B_1\rangle\langle B^2B_1|$) and the SO interaction Hamiltonian:
\begin{equation}
    H_\mathrm{tot} = H_\mathrm{el} + H_\mathrm{so},
\end{equation}
here the SO coupling Hamiltonian is $H_\mathrm{SO} = A_\mathrm{so}L_aS_a$, $L_a$ and $S_a$ are the projection operators of the orbital and spin angular momentum onto the principle axis $a$, respectively. As written in the basis of the electronic states $\{|A^2B_2\rangle, |B^2B_1\rangle\}$, $H_{tot}$ reads~\cite{liu2018rotational}:
\begin{equation}
\label{splitting}
    H_\mathrm{tot} = \begin{pmatrix}E_A&A_\mathrm{so}\Sigma\\
    A_\mathrm{so}\Sigma&E_B
    \end{pmatrix} = \frac{E_B+E_A}{2}+\frac{1}{2}\begin{pmatrix}-\Delta E_0&A_\mathrm{so}\\
    A_\mathrm{so}&\Delta E_0
    \end{pmatrix},
\end{equation}
where $\Delta E_0=E_B-E_A$ is the energy separation between the $|A^2B_2\rangle$ and $|B^2B_1\rangle$ states without the SO interaction, and it could be regarded as the splitting caused by the molecular asymmetry. $\Sigma$ is the projection quantum number of electronic spin onto the principle axis $a$, and we take $\Sigma = \frac{1}{2}$ since its sign does not affect the energy levels. According to Eq.(\ref{splitting}), the energy difference between the two eigenstates of $H_\mathrm{tot}$, $\widetilde{A}$ and $\widetilde{B}$, could be calculated by the following equation:
\begin{equation}
    \Delta E = E_{\widetilde{B}}-E_{\widetilde{A}} = \sqrt{(\Delta E_0)^2+A_\mathrm{so}^2}.
\end{equation}
For SrOPh, the total energy gap is $\Delta E = 305$ cm$^{-1}$, which then gives $\Delta E_0 = 138$ cm$^{-1} \ll A_\mathrm{so}$. This implies that the orbital angular momentum is partially preserved by the strong SO interaction in the SrOPh $\widetilde{A}$ ($\widetilde{B}$) state. Therefore, the SrOPh $\widetilde{A}$ ($\widetilde{B}$) state is similar to a $^2\Pi_{|\Omega|=\frac{1}{2}}$($^2\Pi_{|\Omega|=\frac{3}{2}}$) state in the symmetric top approximation. Noticed that $\Delta E$ is also much larger than $\Delta E_0$ in the other species studied in this work, the SO interaction is expected to dominate the separation between the $\widetilde{A}$ and $\widetilde{B}$ states of all these species.

\newpage
\renewcommand*{\thetable}{S\arabic{table}}
\setcounter{table}{0}
\begin{table*}
    \centering
    \setlength{\tabcolsep}{7pt}
    \renewcommand{\arraystretch}{1.2}
    \begin{tabular}{c c c c c c c}
    \hline
    \hline
    \multirow{2}{*}{Vib. modes } &\multicolumn{2}{c}{SrOPh} && \multirow{2}{*}{Vib. modes } &\multicolumn{2}{c}{SrOPh-3-CF$_3$}\\
    \cline{2-3}\cline{6-7} & Exp.& Theo.& & & Exp.& Theo.    \\
    \cline{1-3}\cline{5-7}
    $\nu_2$ & 54(2) & 54 & &$\nu_2$ & 42(5) & 39\\
    $2\nu_2$ & 102(2) & 108 &  & $\nu_5$&178(5)  &180 \\
    $\nu_3$ & 238(2) & 241 & &$\nu_6$& 219(2) & 222  \\
    $\nu_2\nu_3$ & 297(2) & 294 & & & & \\
    \hline
  \multirow{2}{*}{Vib. modes } &\multicolumn{2}{c}{SrOPh-3-F} && \multirow{2}{*}{Vib. modes } &\multicolumn{2}{c}{SrOPh-3-CH$_3$}\\
    \cline{2-3}\cline{6-7} & Exp.& Theo.&&& Exp. & Theo.\\
    \cline{1-3}\cline{5-7}
    $\nu_1$  & 56(5) & 56 & &$\nu_2$ & 43(5) & 50\\
    $\nu_3$ & 226(2) & 222 & & $\nu_5$& 226(2) &230\\
    \hline
      \multirow{2}{*}{Vib. modes } &\multicolumn{2}{c}{SrOPh-3,4,5-F} && \multirow{2}{*}{ } &\multicolumn{2}{c}{}\\
    \cline{2-3} & Exp.& Theo.&&&&    \\
    \cline{1-3}
    $\nu_2$ & 47(6) & 45 & & &  & \\
    $\nu_4$ & 203(2) & 204 & & &  & \\
    \hline
    \hline
    \end{tabular}
    \caption{Comparison of the observed and calculated frequencies for resolved fundamental vibrational modes of all species studied in this work. Values are given in units of cm$^{-1}$. }
    \label{table:vib.freq.}
\end{table*}

\begin{table*}
    \centering
    \begin{tabular}{c c c c c}
    \hline
        \multirow{2}{*}{Modes} & \multicolumn{4}{c}{SrOPh-CH$_3$}\\ \cline{2-5}
        ~ & Exp.(A) & Theo.(A) & Exp.(B) & Theo.(B)  \\ \hline
        0 & 0.946(6) & 0.926 & 0.918(9) & 0.943  \\ 
        $\nu_2$ & ~ & $7\times10^{-4}$ & 0.011(2) & $2\times10^{-4}$  \\ 
        $\nu_5$ & 0.054(6) & 0.059 & 0.072(9) & 0.044  \\\hline
        \multirow{2}{*}{Modes} & \multicolumn{4}{c}{SrOPh}\\ \cline{2-5}
        ~ & Exp.(A) & Theo.(A) & Exp.(B) & Theo.(B)  \\ \hline
        0 & 0.845(7) & 0.928 & 0.885(5) & 0.945  \\ 
        $\nu_2$ & 0.029(3) & $<10^{-4}$ & 0.054(3) & $<10^{-4}$  \\ 
        $2\nu_2$ & 0.025(3) & $<10^{-4}$ &   &   \\ 
        $\nu_3$ & 0.079(5) & 0.059 & 0.060(3) & 0.043  \\ 
        $\nu_2\nu_3$ & 0.021(3) & $<10^{-4}$ & ~ &   \\ \hline
        \multirow{2}{*}{Modes} & \multicolumn{4}{c}{SrOPh-3-F}\\ \cline{2-5}
        ~ & Exp.(A) & Theo.(A) & Exp.(B) & Theo.(B)  \\ \hline
        0 & 0.956(13) & 0.936 & 0.965(3) & 0.954  \\ 
        $\nu_1$ & 0.007(3) & $9\times10^{-4}$ &  &   \\ 
        $\nu_3$ & 0.037(16) & 0.051 & 0.035(3) & 0.037  \\ \hline
        \multirow{2}{*}{Modes} & \multicolumn{4}{c}{SrOPh-CF$_3$}\\ \cline{2-5}
        ~ & Exp.(A) & Theo.(A) & Exp.(B) & Theo.(B)  \\ \hline
        0 & 0.893(5) & 0.939 & 0.867(11) & 0.950  \\ 
        $\nu_2$ & 0.014(2) & 0.003 & 0.021(5) &  $<10^{-4}$ \\ 
        $\nu_5$ & 0.018(2) & 0.007 & 0.026(6) & 0.007  \\ 
        $\nu_6$ & 0.074(3) & 0.037 & 0.086(8) & 0.035  \\ \hline
        \multirow{2}{*}{Modes} & \multicolumn{4}{c}{SrOPh-3,4,5-F}\\ \cline{2-5}
        ~ & Exp.(A) & Theo.(A) & Exp.(B) & Theo.(B)  \\ \hline
        0 & 0.946(7) & 0.945 & 0.964(4) & 0.963  \\ 
        $\nu_1$ & 0.005(2) & $<10^{-4}$ & 0.008(2) &  $<10^{-4}$  \\ 
        $\nu_4$ & 0.049(7) & 0.044 & 0.028(4) &  0.030 \\ \hline
    \end{tabular}
    \caption{The intensity ratios of all observed vibrational decays of all molecules. The errors indicate the statistical uncertainties from the Gaussian fits. The theoretical VBRs are also added for comparison. }
    \label{table:modes-vbr}
\end{table*}

\begin{table*}
    \centering
    \begin{tabular}{c c}
    \hline
    \hline
       VBRs measurement error source  &  Percentage \\
     \hline
      Contributions from unobserved peaks & $1.1\% - 3.2\%$ \\
      Signal fluctuation & $1.0\%$\\
      Instrument wavelength response & $1.0\%$\\
      Diagonal excitation & $0.5\%$\\
    \hline
      Total error & $1.9\% - 3.5\%$\\
      \hline
     
    \end{tabular}
    \caption{Systematic error budget for the vibrational branching ratio measurements.}
    \label{table:syserr}
\end{table*}

\begin{table*}
    \centering
    \begin{tabular}{c c  c c c c}
    \hline\hline
    \multicolumn{6}{c}{$\widetilde A \rightarrow \widetilde X$ Transition}\\
    
    \hline
        Molecules & Measured & Scaled  & Scaled exluding & Upper & Lower  \\
        &VBR  $S_0$ &  VBR $S'_0$ & main peak $S''_0$ &error bar&  error bar \\      \hline
        SrOPh-3-CH$_3$ & 0.946(6)& 0.933 & 0.933 & 0.015 & 0.006  \\ 
        SrOPh & 0.845(7)& 0.822 & 0.804 & 0.024 & 0.020  \\ 
        SrOPh-3-F & 0.956(13)& 0.943 & 0.941 & 0.019 & 0.013  \\ 
        SrOPh-3-CF$_3$ & 0.893(5) & 0.868 & 0.857 & 0.026 & 0.012  \\ 
        SrOPh-3,4,5-F & 0.946(7)& 0.930 & 0.924 & 0.017 & 0.009  \\ \hline\hline
        \multicolumn{6}{c}{$\widetilde B \rightarrow \widetilde X$ Transition}\\
    
    \hline
        Molecules & Measured & Scaled  & Scaled exluding & Upper & Lower \\
        &VBR  $S_0$& VBR $S'_0$ & main peak $S''_0$ & error bar & error bar \\      \hline
        SrOPh-3-CH$_3$ & 0.918(9) & 0.896 & 0.880 & 0.023 & 0.019  \\ 
        SrOPh & 0.885(5) & 0.869 & 0.859 & 0.017 & 0.012  \\ 
        SrOPh-3-F & 0.965(3)& 0.958 & 0.958 & 0.008 & 0.003  \\ 
        SrOPh-3-CF$_3$ & 0.867(11)& 0.853 & 0.840 & 0.018 & 0.017  \\ 
        SrOPh-3,4,5-F & 0.964(4) & 0.953 & 0.950 & 0.011 & 0.005  \\\hline
    \end{tabular}
    \caption{Measured intensity ratios and scaled VBRs of the diagonal 0-0 decay of all molecules. The scaling process considering contributions of unobserved vibrational decays is detailed in the section of error analysis of VBRs.
    }
    \label{DiagonalVBRs}
\end{table*}

\renewcommand*{\thefigure}{S\arabic{figure}}
\setcounter{figure}{0} 
\begin{figure*}
    \centering
    \includegraphics{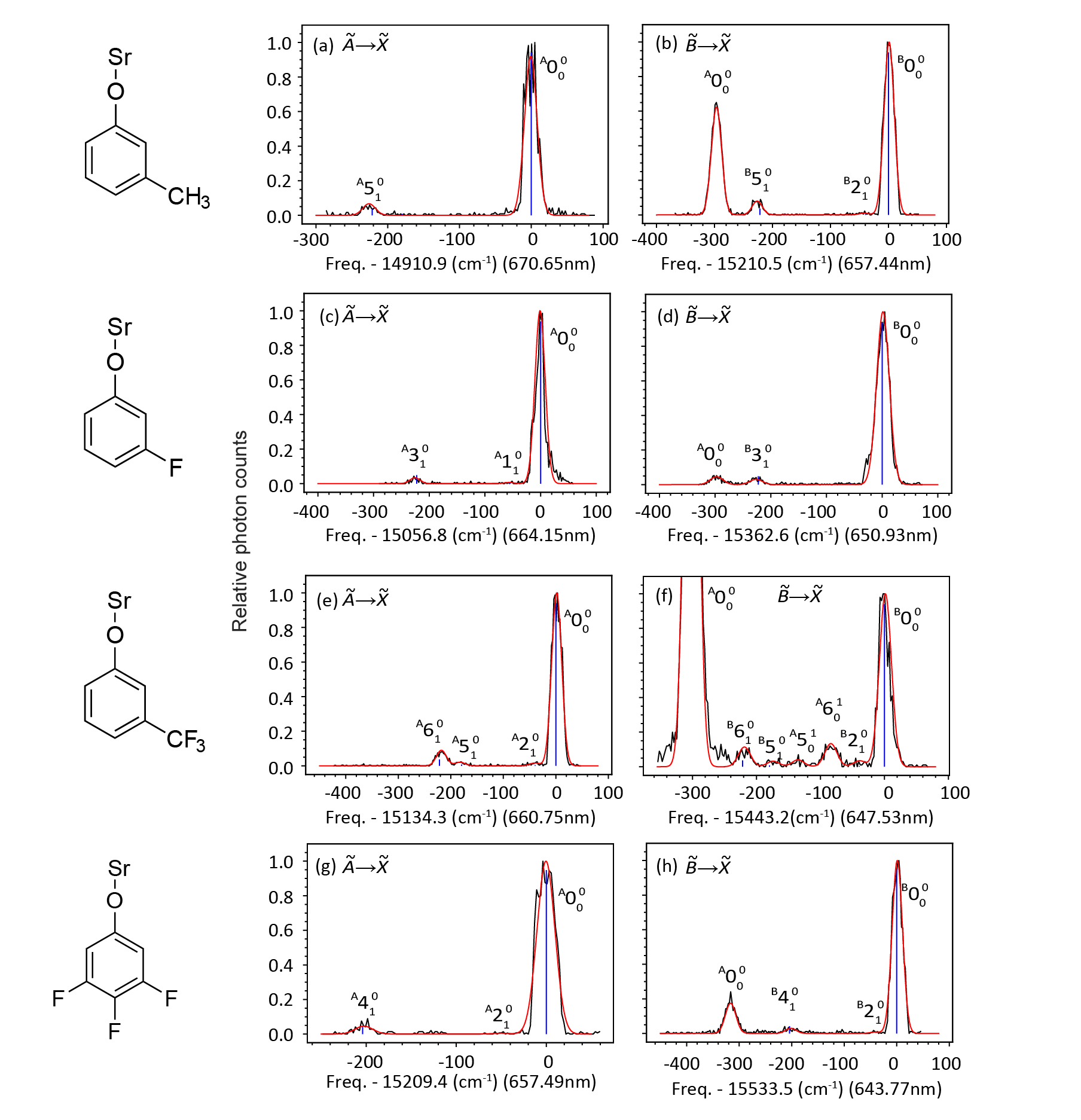}
    \caption{Dispersed fluorescence spectra of all species. The experimental curves (black) are fitted with the Gaussian functions (red). The blue sticks illustrate the vibrational branching ratios of different vibrational modes. The assignments of resolved vibrational peaks are also given. }
    \label{fig:others-dlif}
\end{figure*}

\begin{figure*}
    \centering
    \includegraphics{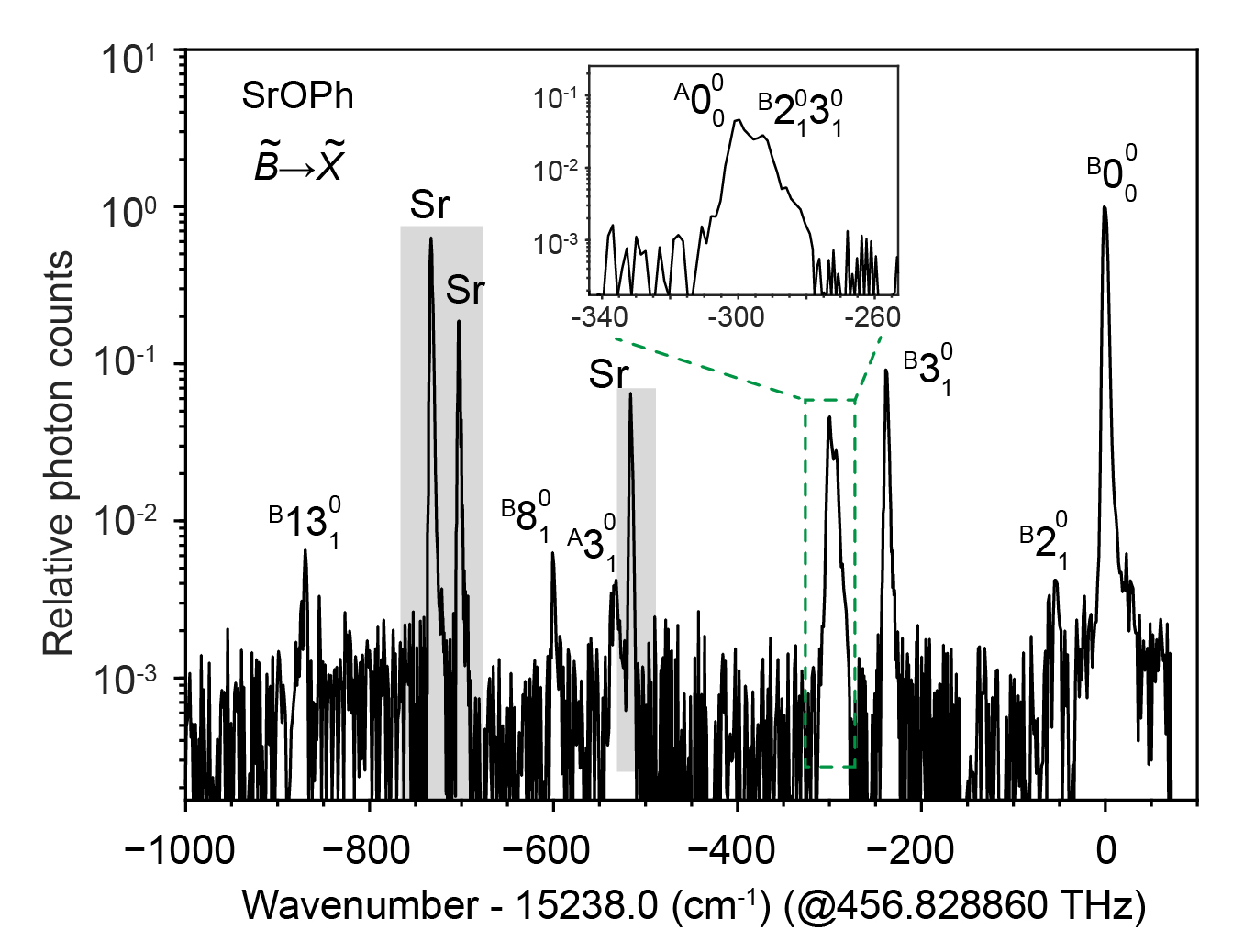}
    \caption{Dispersed spectrum of SrOPh $\widetilde B \rightarrow \widetilde X$ excited by cw laser and measured by a spectrometer coupled with an EMCCD camera. The inset shows the expansion of a broad peak at $-300$ \cm, which is due to the overlapping of two peaks. The assignments of the resolved vibrational peaks are also given.}
    \label{fig:sroph_dlif_cw}
\end{figure*}

\begin{figure*}
    \centering
    \includegraphics{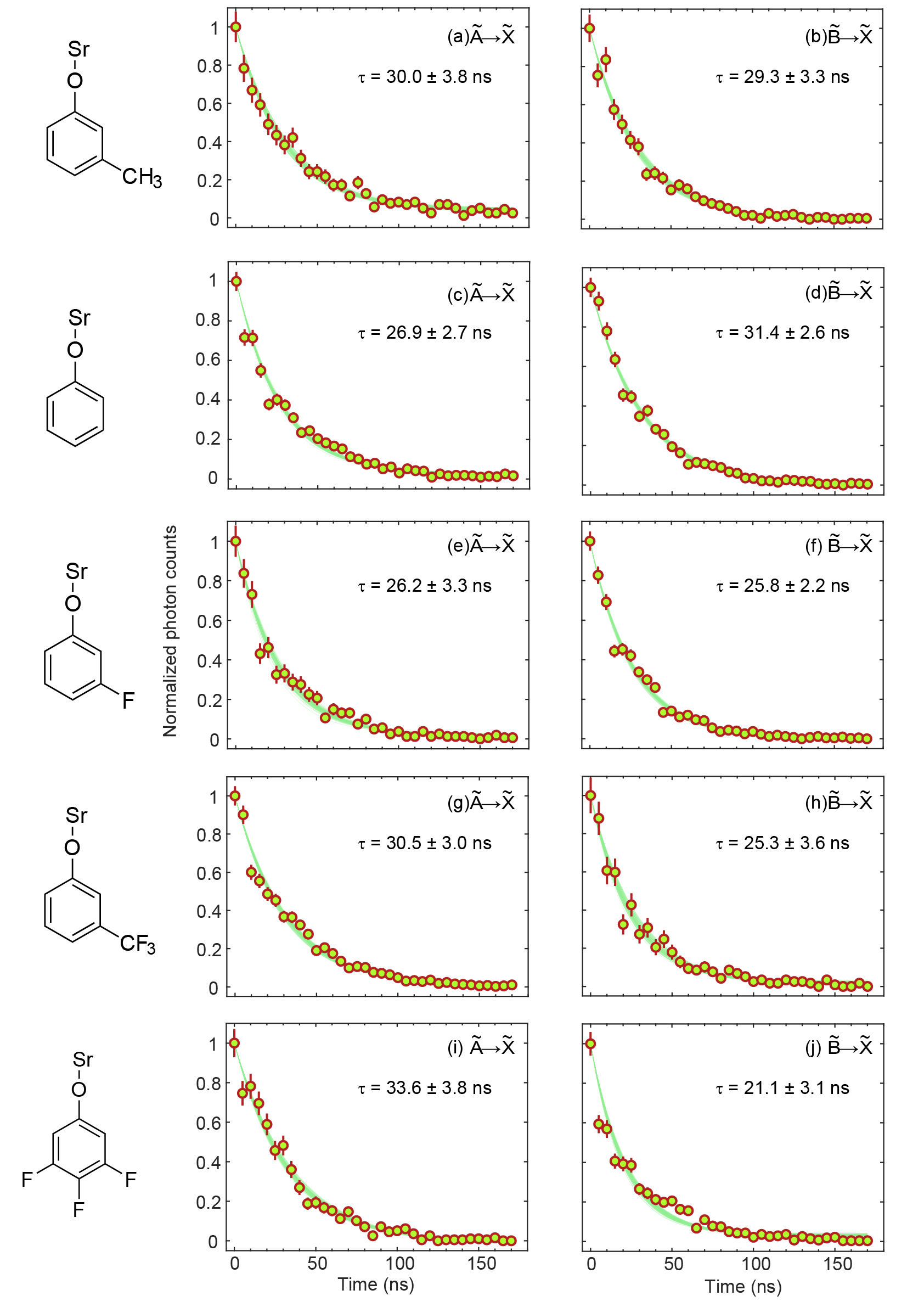}
    \caption{Fluorescence decay traces of all the transitions studied in this work. The experimental data points (red circles) are obtained from the sum of PMT signal for the $0-0$ decay in the DLIF measurements and the respective error bar represents the standard errors. For each trace, the data points are normalized to the maximum signal counts. The radiative lifetimes $\tau$ and errors for all species are estimated from exponential fits (green curves) by bootstrapping the data. 
    }
    \label{fig:lifetime}
\end{figure*}

\begin{figure*}
    \centering
    \includegraphics{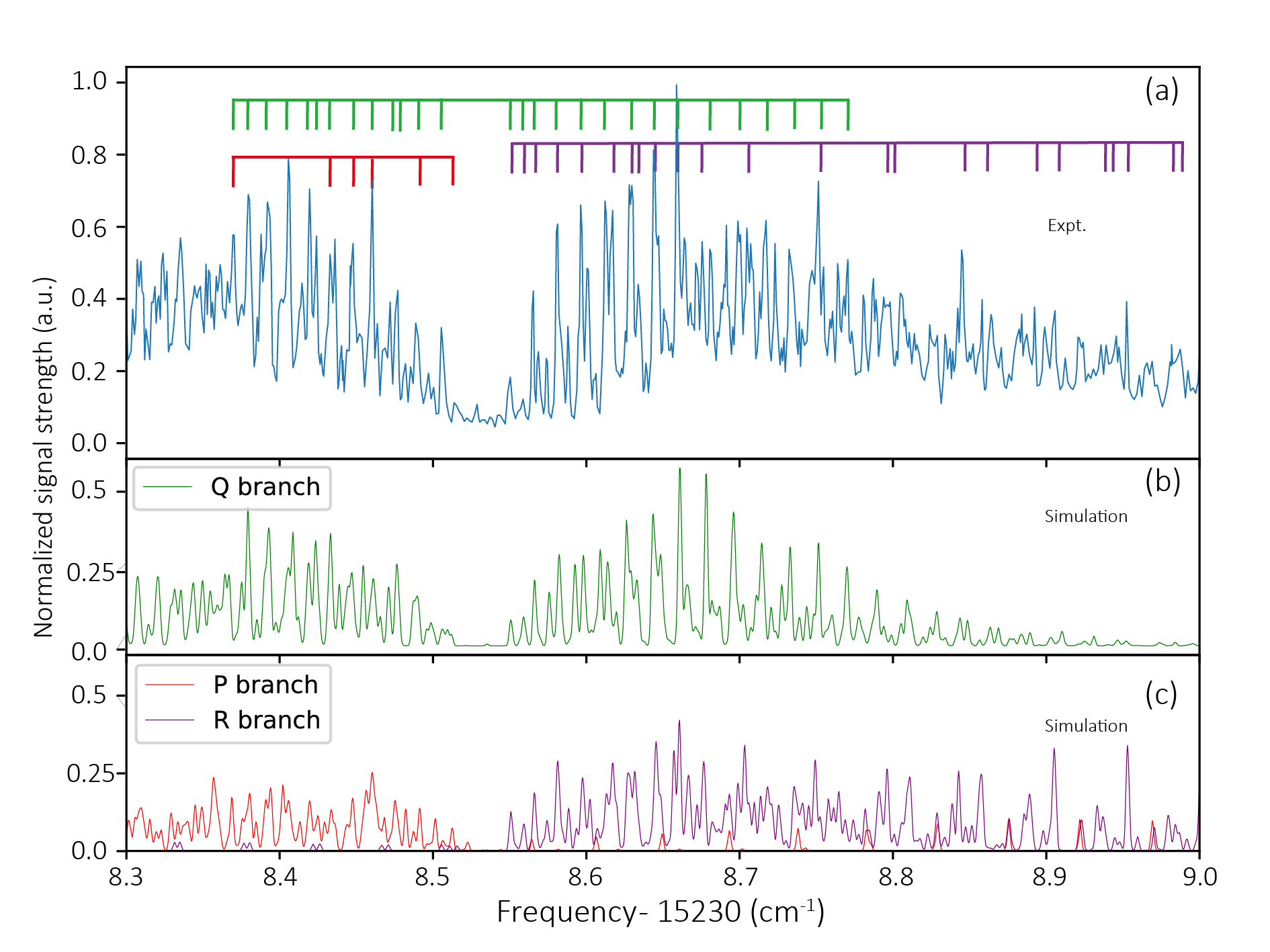}
    \caption{A typical line assignment  near 0-0 transition of SrOPh. The experimental (a) and simulated data (b,c) are normalized to the maximum signal strength in experiment and simulation, respectively. The observed peaks labeled with green, purple and red ticks in the (a) measured spectrum are assigned to the simulation peaks of (b) Q, (c) R and P branch transitions, respectively. The Gaussian linewidth for the simulation is set as 70 MHz to roughly fit the contour, with the rotational temperature set as $T_\mathrm{sim} = 2.5$ K. Each simulation peak in this range usually contains multiple rotational lines, to avoid overfitting, only the strongest $2$ - $4$ lines in each peak are assigned to the corresponding observed peak.}
    \label{SrOPh-cw_assignment}
\end{figure*}

\end{document}